# Measures of Fault Tolerance in Distributed Simulated Annealing

Aaditya Prakash[#1]

[#]*Infosys Limited, India*
[1]`aaadityaprakash@gmail.com`

*Abstract—* In this paper, we examine the different measures of Fault Tolerance in a Distributed Simulated Annealing process. Optimization by Simulated Annealing on a distributed system is prone to various sources of failure. We analyse simulated annealing algorithm, its architecture in distributed platform and potential sources of failures. We examine the behaviour of tolerant distributed system for optimization task. We present possible methods to overcome the failures and achieve fault tolerance for the distributed simulated annealing process.

We also examine the implementation of Simulated Annealing in MapReduce system and possible ways to prevent failures in reaching the global optima. This paper will be beneficial to those who are interested in implementing a large scale distributed simulated annealing optimization problem of industrial or academic interest. We recommend hybrid tolerance technique to optimize the trade-off between efficiency and availability.

## I. INTRODUCTION

Optimization is one problem common to all sciences. There are various methods and algorithms to find the optima. While most methods are susceptible to local optima, Simulated Annealing doesn't suffer from this lacuna due to its probabilistic search space implementation.

With advent of cloud computing technologies Parallel and distributed computing has received revival and most of optimization techniques are now being ported to distributed computing. Simulated Annealing has major advantage in being parallelised but it is not secluded from problems of fault. Due to its stochastic heuristics there are unique faults and thus requires specialized considering for making the system fault tolerance

## II. SIMULATED ANNEALING

Simulated Annealing is one of the most common optimization technique proposed by Kierkpatrick et al[1] in 1983. It is a probabilistic and meta-heuristic algorithm which utilises Metropolis-Hastings to generate its sample search space. As in annealing in metallurgy this technique allows for a global parameter T (Temperature) which is gradually decreased. When T is high any potential solution is accepted, allowing for uphill movement but as T lowers only solution with lower optima are considered. This is controlled by the energy function which follows Boltzmann Probability Distribution.

$$P(E) = e^{-E/kT} \qquad \text{--- (I)}$$

where,
$P(E)$ = Energy Function
$T$ = Temperature
$k$ = Boltzmann constant

As evident from equation(I), at higher T system has uniform probability of being at any state but as T decreases, probability of being at higher energy decreases. Thus by controlling T convergence of algorithm can be controlled.

### A. *Distributed Simulated Annealing*

Distributed Simulated Annealing (DSA) is a variant of simulated annealing optimized for distributed computing. Since the classical Simulated Annealing doesn't store the current best optima[1], the complete process of optimization is thus independent of the past result. This allows for distribution of the search process where all nodes are governed only by one criterion, gradually decreasing Temperature.

As proposed by Arshad et al[2] DSA is not just allowing multiple agents to search the same sample space but having a confluence of the agents to prevent searching of similar nodes even though they are partially governed by stochastic parameters. The algorithm as proposed by them has been discussed in the section III.

Implementation of Distributed Simulated Annealing by Krishnan et al[3] to solve a NP-Hard

Problem of Job Shop scheduling with excellent results are validation towards effectiveness of Simulated Annealing in a parallel system.

### B. Distribution Using MapReduce

Works of Google and Yahoo have made a huge impact on how distributed computing is practiced today. MapReduce with advantages of shared memory, message passing and scalability is just the right tool for parallel implementation of Simulated Annealing, or any other optimization techniques. A recent paper by Radenski[4] on how to implement Distribued Simulated Annealing on MapReduce is a sign that this vista is not only appealing but has already garnered some enthusiasm.

## III. ALGORITHM FOR DISTRIBUTED SIMULATED ANNEALING

All studies of fault tolerance should focus on the points of failure. Algorithms are the sources for design faults, which generally lead to transient faults which are more difficult to overcome than permanent faults. Thus we dissect the algorithm of DSA to point several measures of fault sources.

The algorithm is modified to accommodate the distributed structure. As can be seen from fig 1, the structure of algorithm is highly scalable for multiple processing agents and convergence of solution is much superior than centralized processing. Whenever a given node finds a better minma, it is asynchronously broadcasted to all the working nodes, this allows for faster convergence because this prevents other nodes from discovering same local optima[2].

$P(E_0)$ and $T_{LOW}$ are two user defined parameters and are set according to the specific requirements. If accuracy is more important than the speed then $P(E_0)$ and $T_{LOW}$ are set to relatively high and low values respectively.

## IV. SOURCES OF FAULT IN DSA

### A. Design Faults

Unlike Tabu Search, DSA doesn't have memory structure to store the visited solution [5]. This presents a critical problem with DSA that all the nodes could possibly be stuck in same local optima. This is a critical design fault and could degrade the quality of the output.

Consider a case of search space with global optima in a highly narrow search space with local optima in broad base [Fig 2]. This will with high probability lead to all solutions fixing at local optima.

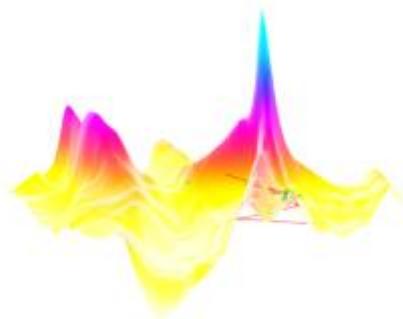

Fig 2: Skewed Global Optima

Another source of design fault is the selection of random number generation for the purpose of initialization of Temperature (T). If the solution space is relatively small then any standard Pseudo

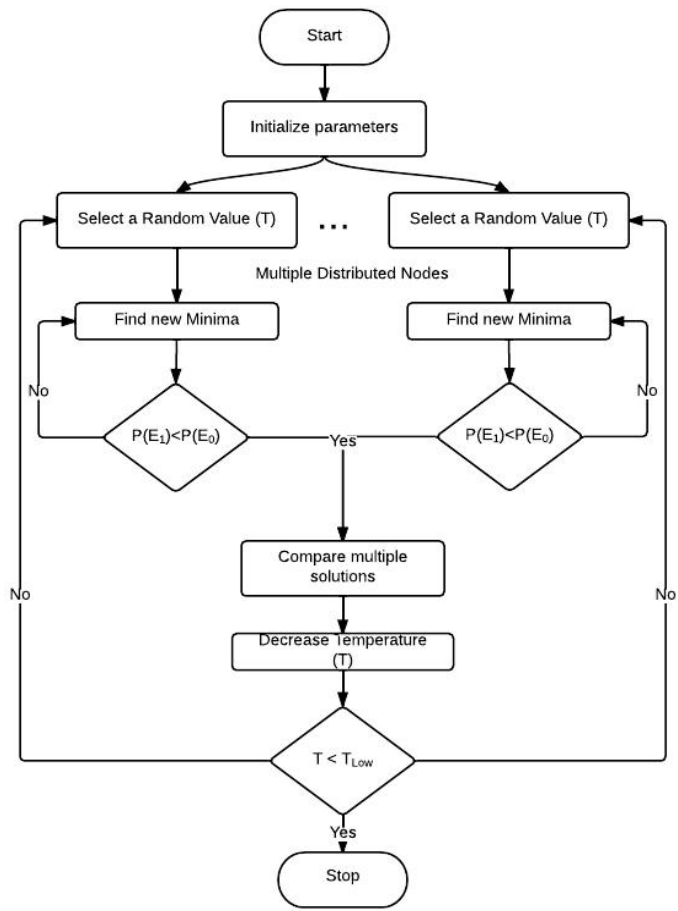

Fig 1: Algorithm for DSA

Random Number Generator (PRNG) should suffice. In the case of larger solution space, probability of converging is dependent on the correctness of PRNGs [7].

*B. Operational Faults*

Implementing any form of distributed computing has its own sources of operational point of failures. While most of the faults concerning the loss of a node or loss of data can be prevented using MapReduce [6] but doing so hampers the effectiveness of the system by compromising on the number active nodes available for the optimization problem.

In Large Scale Distributed Systems, the sources of operational faults are generally due to [8]:
- Independent Failure
- Unreliable Communication
- Insecure Communication
- Costly Communication

While Independent Failure are not of concern considering that the DSA algorithm has been developed to overcome node failure, unreliable communication and insecure communication could lead to unreliable or inaccurate results.

Costly Communication could possibly lead to poor performance. If the overhead of communication of nodes exceed the ratio of fraction of work to total Speedup *(as in Amdahl's Law)* then benefits of distribution of optimization is highly compromised[9].

While DSA is immune to cascading failure, due to absence of atomic multicast repeated execution leads to reduction in performance. A more flexible and hybrid approach to distribution will improve the performance, as documented by Ganeshan et al [10].

## V. TOLERANCE AND RECOVERY

While there have been few studies on study of fault tolerance for a generic distributed systems, only handful of such studies have focused on distributed optimization and more specifically Simulated Annealing.

As presented by Marin et al [11], the flexible adaptive fault tolerant systems for distributed agents have clear advantage over any strategic fault tolerant systems.

DSA is very unique in this regard, since its fault have very difficult detection scheme. Since the solution space is searched in random order, any form of pooling of search space is futile. In this regard MapReduce is highly effective in its implementation of Hash Table to store the intermediate results. While this doesn't allow to confirm if solution space has been searched but allows for quick verification of all intermediate solutions. As shown by Ganjisaffar et al [12] proper tuning of MapReduce to solve optimization problems could lead to improved performance. They have reported AUC above 90% for most of the algorithms tested on MapReduce.

Like in all distributed optimization problems it is not in best interest to have a cold or warm Standby system [10], a mechanism for Hot Standby, where in the existing working node is made to replace the dead node, could be implemented. This replacement will certainly not be able to perform backward error recovery. For Simulated Annealing this is not only prodigal but also unnecessary [1]. Provided that the dead node was not already in best solution i.e global optima, the next search sequence is as good as any other especially when Temperature(T) is still relatively high (equation I).

Therefore a hybrid replication mechanism could be implemented wherein during the high T there will not be any provision for result replication or broadcast, since higher T in Simulated Annealing means lower chances that the system is in low energy state (analogous to energy state of electrons in metallurgical annealing). When T approaches close to the $T_{LOW}$ (T Lower bound), then some of the searching nodes should be replaced as reciprocating nodes. This will ensure that when a solution is found (higher probability in lower T), any failure of the node will not have adverse affect and any intermediate solution will be available to all working agents.

To overcome faults due to unreliable communication channel, the central node (Reducer in the case of MapReduce), should follow the gradient of the slope of Energy (P(E)) and incase of anomalous node, prevent that node from further processing or reset the node's Temperature. This will prevent eccentric nodes from causing undesired deviations in the search space.

## VI. CONCLUSION

Simulated Annealing is one of the most effective optimization techniques to solve for global optima. Of all optimization technique Simulated Annealing has advantage in having independent search iterations thus allowing effective parallelization and distribution. Distributed Simulated Annealing could easily be implemented in traditional parallel systems or in a MapReduce system. Like other distributed systems it has few common sources of fault and few of its own. These faults could lead to either incorrect solution or greatly compromised performance. We have presented potential such sources of fault, their effect and possible remedy to those failures.